\documentstyle[preprint,aps,eqsecnum]{revtex}
\begin{document}

\title {Effects of two dimensional plasmons on
the tunneling density of states}
\author {D. V. Khveshchenko$^1$ and Michael Reizer$^2$}
\address{$^1$ NORDITA, Blegdamsvej 17, Copenhagen DK-2100, Denmark\\
$^2$ Department of Physics, the Ohio State University, Columbus,
Ohio 43210-1106}
\maketitle

\begin{abstract}
\noindent

We show that gapless plasmons lead to a universal $(\delta\nu(\epsilon)/\nu\propto
|\epsilon|/E_F)$
correction to the tunneling density of
states of a clean two dimensional 
Coulomb interacting electron gas. We also discuss a counterpart of this effect in the "composite fermion metal" which forms in the presence of a quantizing perpendicular magnetic field corresponding to the half-filled Landau level. 
We argue that the latter phenomenon might be relevant
for deviations from a simple scaling 
observed by A.Chang {\it et al} in the tunneling $I-V$ characteristics of Quantum Hall liquids.
\end{abstract}

\pagebreak

The phenomenon of suppression of electron tunneling into interacting conductors, known as "zero-bias anomaly", still remains in the center of current theoretical studies.   

This experimentally well-documented phenomenon received its first explanation in the theory of 
the electron density of states (DOS) in Coulomb interacting disordered 
metals \cite{AA}. This theory, however, was formulated in the diffusive regime
and therefore limited to the range of energies $\epsilon$ or, correspondently, bias voltages $V$ small compared to the impurity scattering rate: $\epsilon, V  <  1/\tau$.
 
Recently, an attempt was made to extend the theory
of Ref.\cite{AA} beyond the diffusive regime \cite{RAG}. The authors of Ref.\cite{RAG} found a universal (independent of the strength of Coulomb coupling) correction to the two-dimensional (2D)
DOS: $\delta\nu(\epsilon)/\nu\propto -(E_F\tau)^{-1}[\ln (\epsilon/\Delta)]^2$
in the regime $1/\tau<<\epsilon<<\Delta$,
where the characteristic energy scale $\Delta=v_F\kappa$ is determined by the Debye
screening wavevector $\kappa=2\pi e^2\nu$ proportional to the bare two-spin DOS $\nu=m/\pi$. 
Also, on the basis of the  
calculations performed in \cite{RAG}, a modification of the 
diffusive correction obtained in \cite{AA} was proposed.
Later the same authors generalized the theory of Ref.\cite{AA,RAG} onto the case of non-quantizing
magnetic fields \cite{RAG2}. 

In the present communication we show that the tunneling DOS of a clean 2D Coulomb conductor  
also contains another universal term $\delta\nu(\epsilon)/\nu\propto |\epsilon|/E_F$ which 
is completely independent of impurity scattering and may well become dominant in the ballistic regime. 
With this new term included, 
the tunneling conductance $G(V)$ acquires a linear cusp-like
universal contribution $\delta G(V)/G_0\propto |V|/E_F$.

Unlike Ref.\cite{RAG} where Matsubara technique was used,
we employ the real-time formalism
in order to avoid problems with a somewhat intricate procedure of  
analytical continuation from discrete imaginary frequencies.
The two-spin tunneling DOS is defined as
$$\nu(\epsilon, T)=-{2\over \pi}\int{d^2{\bf p}\over (2\pi)^2}{\rm Im}[{\cal G}^R(\epsilon,{\bf p})]     
                                                                   \eqno(1)$$
In the presence of impurities the non-interacting 
electron Green function has the standard form   
$${\cal G}^R_0(\epsilon, {\bf p})=[{\cal G}^A_0(\epsilon,{\bf p})]^*={1\over \epsilon-\xi_p+i/2\tau},\ \ \ \xi_p={p^2-p^2_F
\over 2m}                            \eqno(2)$$
The interaction correction to DOS is simply related to the electron self-energy 
$$\delta\nu(\epsilon, T)=-{2\over \pi}{\rm Im}\int{d^2p\over (2\pi)^2}[{\cal G}^R_0(\epsilon,{\bf p})]^2
\Sigma^R(\epsilon,{\bf p})                                                 \eqno(3)$$
 In the quasi-ballistic regime of large momentum and energy transfers
($1/\tau < v_Fq, \omega$) the self-energy $\Sigma(\epsilon, {\bf p})$ is
given by the expression
$$ \Sigma^R(\epsilon,{\bf p})=\int{d\omega d^2{\bf q}\over (2\pi)^3}\biggl
[{\rm Im}{\cal G}_0^A(\epsilon+\omega,{\bf p}+{\bf q})
V^A({ q})\tanh\biggl({\epsilon+\omega\over 2T}\biggr)$$
$$+{\cal G}^R_0(\epsilon+\omega,{\bf p}+{\bf q})
{\rm Im}V^A(\omega,{q})\coth\biggl({\omega\over 2T}\biggr)\biggr]                                       \eqno(4)$$
where $V^A(\omega,{ q})$ is a dynamically screened 2D Coulomb potential  
$$V^A(\omega,q)={V_0(q)\over 1+V_0(q)P^A(\omega, q)},\\\ V_0(q)=2\pi e^2/q     \eqno(5)$$
In this formula   $P^A(\omega,q)$ is the impurity-dressed 
polarization operator  
$$P^A(\omega,q)=
{\nu}\biggl(1-{\Gamma(\omega,q)(\omega-{q^2/2m})\over \sqrt{(\omega-{q^2/2m}+i/\tau)^2-v_F^2q^2}}\biggr)                                                                           \eqno(6)$$
which also includes the impurity vertex correction
$$
\Gamma(\omega,q)=\biggl(1-{i/\tau\over 
\sqrt{(\omega-{q^2/2m}+i/\tau)^2-v_F^2q^2}}\biggr)^{-1}
\eqno(7)$$
After the integration in (3) over the electronic momentum  
one arrives at the expression
$${\delta\nu(\epsilon, T)\over \nu}={1\over (2\pi)^2}\int d\omega 
\tanh\biggl({\epsilon+\omega\over 2T}\biggr){\rm Im} 
\int_0^\infty dqqV^A(\omega,q){[\Gamma(\omega,q)]^2(\omega-{q^2/2m}+i/\tau)\over 
[(\omega-{q^2/2m}+{i/\tau})^2-v^2_Fq^2]^{3/2}}        \eqno(8)$$
A straightforward analysis of Eq.(8) shows that the 
range of transferred energies and momenta $1/\tau < \omega < v_Fq$, where the Coulomb potential is statically screened ($V(q)={1\over \nu}{\kappa\over q+\kappa}$), yields only a small contribution to DOS.
The $q$-dependence of $V(q)\approx 1/\nu$ is weak, and, with logarithmic accuracy, this contribution coincides with that of
a short-ranged potential $V(q)=V_0$:
$${\delta\nu(\epsilon, T)\over \nu}=-{V_0\nu\over 4\pi\epsilon_F\tau} \ln {\Delta\over {\rm max}\{|\epsilon|, T\}}        \eqno(9)$$
The authors of Ref.\cite{RAG} 
used the coordinate space representation
to demonstrate that this term occurs due to  
the interference between scattering off a single impurity and off
Friedel oscillations of the electronic density caused by the same impurity.

However, the overall $q$-integral in (8) is dominated by the interval of momenta $\omega^2/\Delta <v_Fq <\omega$ where 
the "anti-screened" potential $V(\omega,{\bf q})$ develops a plasmon pole  
at $\omega=v_F(\kappa q/2)^{1/2}$. As we show below, this gapless collective mode plays a role which is somewhat similar to that of a diffusion pole $\omega=iDq^2$
appearing in the disordered regime $\omega, v_Fq  < 1/\tau$. 

The contribution resulting from the above range of momenta can be readily found
$${\delta\nu(\epsilon, T)\over \nu}={1\over (2\pi)^2}{\kappa \over \nu}
\int d\omega \tanh\biggl({\epsilon+\omega\over 2T}\biggr){\rm Im} 
\biggl({\omega+i/\tau\over \omega}
\int_0^{\sim\omega/v_F} {dq\over {(\omega+i/\tau)\omega-q\kappa v_F^2/2}}\biggr)$$
$$=-{2\over (2\pi)^2\nu v^2_F}
\int d\omega \tanh\biggl({\epsilon+\omega\over 2T}\biggr)
{\rm Im}\biggl[{\omega+i/\tau\over \omega}
(\ln{\Delta\over |\omega|}+i\pi {\omega\over |\omega|})\biggr]       \eqno(10)$$
The first term in (10) which stems from the real part of the $q$-integral reproduces  
the correction obtained in Ref.\cite{RAG}:
$${\delta_1\nu(\epsilon, T)\over \nu}=-{2\over (2\pi)^2}{1\over v_F^2\nu\tau}
\int {d\omega\over \omega} 
\tanh\biggl({\epsilon+\omega\over 2T}\biggr)\ln {\Delta\over |\omega|}=-{1\over {4\pi E_F\tau}} \biggl(\ln {\Delta\over {\rm max}\{|\epsilon|, T\}}\biggr)^2          \eqno(11)$$
which appears to be greater than (9) by an extra logarithmic factor.

The second term originating from the imaginary part of the integral over $q$ 
constitutes our new result 
$${\delta_2\nu(\epsilon, T)\over \nu}=-{2\over (2\pi)^2v_F^2\nu}
\int {d\omega}{\omega\over |\omega|} 
[\tanh\biggl({\epsilon+\omega\over 2T}\biggr)$$
$$=-{1\over 2\pi v_F^2\nu}
\int_0^{\sim\Delta} {d\omega}
[\tanh\biggl({\epsilon+\omega\over 2T}\biggr)+\tanh\biggl({\omega-\epsilon\over 2T}\biggr)-2]=
{{\rm max}\{|\epsilon|, 2T\ln 2 \}\over 2E_F}                       \eqno(12)$$
where we subtracted a constant term $\delta\nu(0,0)$ to avoid a divergence at the upper limit.
The new term (12) exceeds (11) in the whole range of energies
$\tau^{-1} (\ln \Delta\tau)^2 < \epsilon < E_F$.

The DOS correction (12) is valid, strictly speaking, 
only at weak coupling where one can take an advantage of the condition
$\kappa<<k_F$. The latter guarantees that 
all relevant momenta $q\sim \omega^2/\Delta v_F< \kappa$ are small compared to $k_F$. 
Hence, at weak coupling one can justify the above neglect of the recoil term $q^2/2m$
in the integrand in Eq.(8).

In the absence of momentum conservation 
the applied voltage bias and/or temperature dependent
correction to the differential tunneling conductance $G(V)=dI/dV$ of a point contact between two identical 2D leads
(e.g., coupled layers in a double-well system)
is simply related to the DOS correction:
$${\delta G(V,T)\over G_0}={d\over dV}\int d\epsilon [n(\epsilon)-n(V-\epsilon)]
\biggl({\delta\nu(\epsilon, T)\over \nu}+{\delta\nu(V-\epsilon,T)\over \nu}\biggr)
={{\rm max}\{|V|, 2T\ln 2 \}\over E_F}     \eqno(13)$$
where the zero-bias value of the conductivity 
$G(0,0)=G_0$ includes the (negative) $V$- and $T$-independent term  subtracted in Eq.(12).

It is worth mentioning that a universal correction to the tunneling conductance similar to (13) was obtained in the case of tunneling through a uniform barrier which imposes an additional condition of
partial momentum conservation \cite{SG}.
The physical origin of this effect is, however, completely different
from ours:
the correction $G(V)/G_0\propto |V|/E_F$ was obtained in \cite{SG} for the case 
of a short-range potential and was shown to be due to Friedel oscillations of the electronic density induced by the barrier.

Alternatively, the correction to $G(V)$ can be found directly by means of the semiclassical method
of the tunneling action developed in \cite{N}.
Employing this formalism, we obtain the tunneling conductivity 
(hereafter we put $T=0$ for simplicity) 
$$
\delta G(V)\propto {\rm Im} \int^{\infty}_{0} {dt\over t} \exp(iS(t)- i V t) 
\eqno(14)$$
in terms of the time-dependent
action of the electrostatic potential excited in the process of tunneling:    
$
S(t)= \int d\omega |J(\omega,t)|^2 S(\omega)$.
Here $J(\omega, t)=(1-e^{i\omega t})/\omega$ is a spectral function of the point contact,
and the kernel 
$$S(\omega)=\int {d^2{\bf q}\over (2\pi)^2} {{\tilde V}^A_0(q)\over {1+P^A(\omega,q){\tilde V}_0(q)}}   \eqno(15)$$
is given in terms of
 the interaction potential ${\tilde V}_0(q)=V_0(q)(1-e^{-qd})$ which accounts for the
screening induced by the other layer in the double-well system with the spacial
separation $d$ between
the layers. 

At weak coupling the exponent in (15) can be expanded
in powers of the action $S(t)$ which approaches a constant value as the charge-spreading time $t\sim 1/V$ tends to infinity at vanishingly small biases 
($S(t)={\rm const} +O(1/t)$). The first order correction to the conductivity
can be cast in the form 
$$
{\delta G(V)\over G_0}={\rm Im}\int^V_0 {d\omega\over \omega^2} S(\omega)   \eqno(16)$$
From Eq.(16) we reproduce the result (13) at $V>{\rm max}\{(v_F\Delta/d)^{1/2},1/\tau\}$
in an agreement with the DOS calculation.
 
Now we are going to show that the non-analytical correction  to DOS due to 2D plasmons 
becomes even stronger in a quantizing magnetic field
corresponding to the half-filled Landau level when the Coulomb interacting 2DEG forms
a compressible metal-like state \cite{HLR}.

On formal grounds, 
this problem appears to be similar to that of the relativistic electron-electron interaction
resulting from the coupling of electrons to transverse photons \cite{R}. 
The previous analysis of the ballistic regime in the 3D relativistic problem revealed singular 
corrections to DOS and other thermodynamic quantities 
\cite{R} which all exhibit deviations from the conventional Fermi liquid behavior. 
In a recent work this analysis 
was extended onto renormalization of the Drude conductivity \cite{KR}. 

Physically, it is a bulk magnetoplasmon with a strongly overdamped dispersion which plays the role of the transverse photon in the gauge theory of the half-filled Landau level \cite{HLR}. 
It is described by the transverse gauge propagator $D^A_{\perp}(\omega,q)$ which is simply
proportional to the density correlation function.
The latter can be readily expressed in terms of the   
physical (in general, frequency- and momentum-dependent) conductivity $\sigma(\omega, q)$
which is, in turn, inversely proportional
to the irreducible quasiparticle conductivity $\sigma_{cf}(\omega, q)$ of a nearly Fermi liquid-like   
"composite fermion metal" \cite{HLR}:
$$\sigma(\omega,q)\approx (e^2/\phi h)^2/\sigma_{cf}(\omega, q)     \eqno(17)$$
Here $\phi$ stands for the number of the magnetic flux quanta bound to each composite fermion 
($\phi=2$ for the half-filled Landau level), and the quasiparticle 
conductivity $\sigma_{cf}(\omega, q)$ is approximately given
by the Fermi liquid expression. In particular,
in the relevant regime of $\omega < qv_F$ one has $\sigma_{cf}(\omega, q)=k_F/4\pi q$
and, hence,  $\sigma(\omega, q)\propto q$. 

Despite the obvious fact that the problem of the half-filled Landau level does not feature any small
parameter, it became customary to consider $\phi$ as such a parameter while expanding  
the quasiparticle self-energy $\Sigma_{cf}(\epsilon, {\bf p})$ in powers
of $\phi$.
Previous studies \cite{AIM} revealed that the interaction mediated by transverse gauge photons
obeys an analogue of the Migdal theorem, according to which higher order corrections to $\Sigma_{cf}(\epsilon, {\bf p})$ do not change the functional form inferred from the lowest order term.
This simplification can be traced back to the condition $v_Fq\sim (\Delta\omega)^{1/2} >> \omega$
(see Eq.(18) below) satisfied in a typical scattering of composite fermions via absorption or emission of gauge photons.

It was also claimed in the course of the studies \cite{AIM} 
that the mean-field Fermi liquid-like quasiparticle DOS $\nu_{cf}(\epsilon)=m_{cf}/2\pi$ remains intact,
since $\Sigma_{cf}(\epsilon, {\bf p})$ is only weakly dependent on momentum $\bf p$ in the vicinity of the quasiparticle Fermi surface.
This conclusion, however, would only hold true provided $\Sigma_{cf}(\epsilon, {\bf p})$
remained independent of $\bf p$ at all momenta.    

In what follows we demonstrate that because of the gauge interactions
$\nu_{cf}(\epsilon)$ does receive
non-analytic non-Fermi-liquid-like corrections which are even stronger than their Coulomb 
counterpart (12).

To calculate the first order 
correction to the quasiparticle DOS we make use of Eqs.(3-5) where the screened
Coulomb interaction $V^A(\omega,q)$ has to be substituted by 
the transverse gauge coupling
$({{\bf p}\times{\bf q}\over mq})^2D^A_{\perp}(\omega,q)$ with
$$D^A_{\perp}(\omega,q)={\phi^2 \sigma(\omega,q)
\over i\omega+ \sigma(\omega,q)q^2V_0(q)}   \eqno(18)$$
After the integration over $\bf p$ we arrive at the expression 
$${\delta\nu_{cf}(\epsilon, T)\over \nu_{cf}}={1\over (2\pi)^2}\int d\omega 
\tanh\biggl({\epsilon+\omega\over 2T}\biggr)
 \int_{0}^{\sim k_F} qdq
{\rm Im}\biggl
[{D^A_{\perp}(\omega,q)\over q^2}({\omega\over (\omega^2-v^2_Fq^2)^{1/2}}-1)\biggr]
  \eqno(19)$$
Subtracting a divergent constant in the same way as in Eq.(11), we obtain
$${\delta\nu_{cf}(\epsilon, T)\over \nu_{cf}}\sim -\int_0^{\sim\Delta} {d\omega\over |\omega|^{1/2}}
\biggl[\tanh\biggl({\epsilon+\omega\over 2T}\biggr)
       +\tanh\biggl({\omega-\epsilon\over 2T}\biggr)-2\biggr]\sim
\biggl({{\rm max}\{|\epsilon|,T\}\over \Delta}\biggr)^{1/2}    \eqno(20)$$
where the energy scale $\Delta$ re-appears as an order-of-magnitude estimate of the 
effective composite fermion Fermi energy
which is solely due to the Coulomb interaction.

Referring to the preceeding discussion, we expect that at $V << \Delta$ the functional behavior
of the  
exact composite fermion DOS is not altered by higher order terms, although 
the overall numerical factor in (20) remains undetermined.
   
For the short-ranged (screened) potential $V_0(q)=V_0$ we obtain an even larger contribution
$${\delta\nu_{cf}(\epsilon, T)\over \nu_{cf}}\sim 
\biggl({{\rm max}\{|\epsilon|,T\}\over \Delta}\biggr)^{1/3}      \eqno(21)$$
in agreement with the conclusion drawn in Ref.\cite{HLR} that in this case 
the fluctuations of electron density are only loosely controlled by the repulsive forces and, therefore, the effects 
of magnetoplasmons become even more pronounced. 
 
In view of the drastic difference of the true
quasiparticle excitations of the compressible "composite fermion metal" 
from the original electrons \cite{HLR}, the above Eq.(20,21) can not be immediately applied to the
analysis of tunneling which involves real electrons rather than
bulk quasiparticles.  

The standard picture of an electron viewed as a quasiparticle bound to $\phi$ magnetic flux quanta \cite{HLR}
implies the following relation between the corresponding DOSs:
$$\nu(\epsilon)=
\int d\epsilon^{\prime}\nu_{cf}(\epsilon^{\prime})
\int dt e^{i(\epsilon -\epsilon^{\prime}) t} \exp{iS(t)}
  \eqno(22)$$
where the leading non-analytic correction to the quasiparticle DOS $\nu_{cf}(\epsilon)$ at $\epsilon << \Delta$ is given by Eqs.(20,21).

The relation (22) can be derived on the basis of the eikonal-type approximation for the tunneling
electron's Green function taken 
at the location of the point-contact: ${\cal G}(t,{\bf 0})\approx {\cal G}_{cf}(t,{\bf 0})\exp iS(t)$.
Although in this approximation, which is certainly valid for narrow constrictions, the electron's semiclassical trajectory is purely time-like, and the tunneling electron can be represented by a static classical charge source: 
$\int d\omega e^{i\omega\tau}J(\omega,t)=e[\theta(\tau -t)-\theta(\tau)]$, yet there is a non-trivial spatial component
of the composite fermion current ${\bf j}_{cf}$, which is due to the total current conservation
(${\bf \nabla}{\bf j}_{cf}+\partial_t J=0$). It is the current
${\bf j}_{cf}$ which couples to the transverse gauge photons 
(magnetoplasmons) and generates 
the tunneling action 
$S(t)={1/2}\int {d\omega d{\bf q}\over (2\pi)^3}
|{\bf j}_{cf}\times {\bf q}|^2{D_{\perp}(\omega, q)/q^2}$. The kernel of this action can be cast in the form similar to (15):
$$S(\omega)=\int {d^2{\bf q}\over (2\pi)^2} {{\tilde V}_0(q)\over {1+i\sigma(\omega, q)q^2{\tilde V}_0(q)/\omega}}   \eqno(23)$$
In contrast to the case of zero field, the exponent in (23) can not be expanded
in powers of $S(t)$ which diverges as $t^{1/2}$ as $t\rightarrow\infty$.
Instead, the integral (23) should be taken by means of the saddle point method.

The leading exponential behavior
of the tunneling conductance $G(V)\propto e^{-(\Delta/|V|)}$
which holds at ${\rm max}\{1/\tau, e^2/d\} << V << \Delta$ (and $G(V)\propto e^{-(\Delta/|V|)^{1/2}}$
at $1/\tau << V << e^2/d$ in the screened regime, which occurs, provided that
$1/\tau << e^2/d$) was found in \cite{HPH}.  In \cite{HPH} the approximation $G_{cf}(t,{\bf 0})\propto 1/t$ was made, which is equivalent to a constant $\nu_{cf}(\epsilon)$. 
In principal, our Eqs.(20,21) can be used to determine subdominant power-law corrections which may become important at large bias voltages.

The correction to the composite fermion DOS might be of an even greater importance in situations where the tunneling conductance undergoes a milder, only a power-law, suppression, which implies a
logarithmic divergence of the tunneling action: $S(t)\propto \ln t$. 
Among those situations are such experimentally
relevant set-ups as in-plane tunneling through a narrow constriction \cite{M}
and tunneling into the edge of a Quantum Hall system \cite{C}. 

Until recently,  most of the theoretical work in the field was focused on tunneling between
edge states associated with  
incompressible Fractional Quantum Hall states in the bulk.
For the primary case of Laughlin liquids at
bulk filling fractions $\nu_b = 1/2n+1$ which are characterized by a single chiral edge mode,
the point-contact conductance was found exactly by means of a mapping
onto the 1D boundary sine-Gordon model with the coupling constant
$g=\nu_b$ \cite{FLS}.  
In the range of bias voltages $T<< V << T_0$, where the temperature scale $T_0$
is an intrinsic characteristic of the point contact, 
the exact solution amounts to a power-law consistent with 
the behavior predicted for a whole class of more general Jain fractions $\nu_b=n/(2mn+1)$ with positive integer $n$ and $m$.  For all these fractions,
which can be described in terms of $n$ edge modes of the same chirality, 
the non-ohmic point-contact conductance
is governed by the $n$-independent universal exponents: $G(V)\propto |V|^{4m}$ \cite{W}.
On the contrary, in the case of Jain fractions with negative $n$ the  
edge modes propagate in opposite directions, and in the absence of equilibration between 
different modes  
$G(V)$ was found to exhibit non-universal exponents.
With such an equilibration included, the conductance was found to 
depend on $|n|$ as $G(V)\propto |V|^{4(m-1/|n|)}$ \cite{KFP}.

Despite the solid theoretical predictions, the detailed fit of the  $\nu_b=1/3$
experimental data from \cite{C} reveals a deviation  
from the simple scaling \cite{CF}. In \cite{CF} this deviation was assigned 
to irrelevant tunneling
operators, which contribute to $G(V)=\sum_{k=1}^{\infty}G_k (|V|/T_0)^{4k}$ as higher order terms
with $k>1$ 
(in order to avoid confusion we emphasize that in contrast to 
the analysis of Ref.\cite{CF} where tunneling into the Quantum Hall edge from a 3D lead was considered,
our discussion refers to the problem of in-plane tunneling through a point contact connecting two identical Quantum Hall liquids).

As an alternative approach, the recent theory \cite{SLH} based on the idea of composite fermions with a constant 
$\nu_{cf}$ facilitated the calculation of the tunneling conductance at arbitrary $\nu_b$. This theory  treats both incompressible and compressible bulk states in the same manner, 
and a generic Quantum Hall liquid in the interval $1/3 < \nu_b < 1$ is described as a system of composite fermions with $\phi=2$ 
exposed to the residual magnetic field $B_{cf}=B(1-2\nu_b)$.
The exponent in the power-law dependence of $G(V)$ found in \cite{SLH} varies 
as a function of conductivity,
compressibility, and other parameters of the "composite fermion
metal", although at Jain fractions it approaches the abovementioned universal values provided
the D.C. conductivity  $\sigma(0,0)$ is reasonably small. 

In case of in-plane tunneling through a point contact immersed into a compressible composite fermion system, one can 
elaborate on the ideas of Ref.\cite{CF} and to perform a mapping onto a parallel series of 
1D boundary sine-Gordon models with (in general, non-universal) coupling constants $g_N$,
which are related to the exponents in the asymptotic
expansion of the exact electronic Green function 
${\cal G}(t,{\bf 0})=\sum_{N=1}^{\infty}c_N t^{-1/g_N}$ at large $t$ \cite{K}.

The contributions of these auxiliary "channels" labeled by $k$ 
sum up to the total conductance 
$$G(V)=\sum_{N=1}^{\infty}\sum_{k=1}^{\infty}G_{Nk}\biggl({|V|\over T_{0,N}}\biggr)^{2k(1/g_N -1)}  \eqno(24)$$ 
where $G_{NK}\propto |c_N|^2$.

To account for the DOS correction (20), it suffices to keep the $N=2$ term with 
$g^{-1}_2=g^{-1}_1+1/2$ in the sum over $N$.
Searching for possible deviations from a simple scaling, one has to make a comparison between
the first two competing terms: $N=2, k=1$ and $N=1, k=2$. Using the 
value of $g_1(\nu_b)$, which was estimated in \cite{SLH}
for filling fractions $\nu_b$ from the interval 
$1/3 < \nu_b < 1$ as $\approx {\rm min}\{2/\nu_b -2, 2\}$,   
we conclude that the former term can dominate over the latter one at all $\nu_b < 4/5$. 
Thus, at large bias voltages the corrections (20,21) may lead to stronger deviations from the 
simple scaling than the terms with $k>1$ resulting from irrelevant tunneling operators. 

From bulk resistivity measurements 
one obtains $1/\tau_{cf}\sim 0.5K$ , whereas, according to the analysis
carried out in \cite{CF},  the data on tunneling into the $\nu_b=1/3$ edge \cite{C} depart from the simple scaling in the range
of bias voltages $10^2 \mu V < V < 10^3\mu V$,
which can be indeed identified as the ballistic regime for the composite fermions. 
        
To conclude, in the present paper we demonstrate that 
gapless 2D plasmons affect the tunneling DOS even in the ballistic regime.
At zero magnetic field we find a new, impurity-independent, non-analytic correction
which leads to the linear cusp-like contribution to the tunneling conductivity.
In the presence of a quantizing magnetic field
such an auxiliary quantity as the DOS of composite fermions is shown to  
receive an even larger, sublinear,  contribution.
We speculate that at large bias voltages the latter effect may cause substantial deviations from 
a simple scaling of the non-ohmic $I-V$ characteristic of a point contact connecting two
Quantum Hall liquids.  

One of the authors (M.Yu.R.) acknowledges support from US Office of
Naval Research.

\end{document}